\documentclass[11pt,draftclsnofoot,onecolumn]{IEEEtran}
\usepackage{amssymb}
\usepackage{amsmath}
\usepackage[dvips]{graphicx,color}
\usepackage{epsfig}
\usepackage{thrmappendix}

\renewcommand{\vec}[1]{\mbox{\boldmath$#1$}}

\newcommand{\thmend}{\hspace*{\fill}~\QEDopen\par\endtrivlist\unskip}
\newcommand{\myproof}[1]{\noindent\hspace{2em}{\itshape #1 }}

\renewcommand*{\today}{%
    \ifcase\month \or January\or February\or March\or April\or May\or June\or
    July\or August\or September\or October\or November\or December\fi
    \space\number\year}

\makeatletter 
\long\def\@makecaption#1#2{%
    \vskip\abovecaptionskip
    \sbox\@tempboxa{#1. #2} 
    \ifdim \wd\@tempboxa >\hsize
        {#1. #2\par} 
    \else
        \global \@minipagefalse
        \hb@xt@\hsize{\hfil\box\@tempboxa\hfil}%
    \fi
    \vskip\belowcaptionskip}
\makeatother 

\renewcommand{\baselinestretch}{0.99}

\begin{document}

\markboth{SUBMITTED TO THE IEEE TRANSACTIONS ON INFORMATION
THEORY}{}

\title{Beam Selection Gain Versus Antenna Selection Gain}

\author{Dongwoon Bai, Saeed S. Ghassemzadeh, Robert R.
Miller, and Vahid Tarokh%
\thanks{This work was presented in part at the IEEE
Vehicular Technology Conference, Calgary, Canada, September 2008.}
\thanks{Dongwoon Bai and Vahid Tarokh are with the School of Engineering and Applied
Sciences, Harvard University, Cambridge, MA 02138, USA (email:
dbai@fas.harvard.edu; vahid@seas.harvard.edu).}
\thanks{Saeed S. Ghassemzadeh and Robert R. Miller
are with AT\&T Labs. -- Research, Florham Park, NJ 07932, USA
(email: saeedg@research.att.com; rrm@research.att.com).}
\thanks{Manuscript submitted to the IEEE Transactions on Information Theory on February 5, 2009.}
}


\newtheorem{theorem}{Theorem}
\newtheorem{lemma}{Lemma}
\newtheorem{corollary}[theorem]{Corollary}

\maketitle                   

\begin{abstract}
We consider beam selection using a fixed beamforming network (FBN)
at a base station with $M$ array antennas. In our setting, a Butler
matrix is deployed at the RF stage to form $M$ beams, and then the
best beam is selected for transmission. We provide the proofs of the
key properties of the noncentral chi-square distribution and the
following properties of the beam selection gain verifying that beam
selection is superior to antenna selection in Rician channels with
any $K$-factors. Furthermore, we find asymptotically tight
stochastic bounds of the beam selection gain, which yield
approximate closed form expressions of the expected selection gain
and the ergodic capacity. Beam selection has the order of growth of
the ergodic capacity $\mathnormal{\Theta}(\log(M))$ regardless of
user location in contrast to $\mathnormal{\Theta}(\log(\log(M)))$
for antenna selection.
\end{abstract}

\section{Introduction}      
\label{sec:intro}           

Deploying multiple antennas at a base station dramatically increases
spectral efficiency. While multiple-input/multiple-output (MIMO)
systems require multiple RF chains and elaborate signal processing
units, \textit{Antenna selection} has been an attractive solution
for multiple antenna systems because only one RF chain is required
to use the antenna with the highest signal-to-noise ratio (SNR).

With promise of higher spectral efficiency, we focus on \textit{beam
selection} instead of antenna selection using a FBN at a base
station which deploys $M$ multiple linear equally spaced
omnidirectional array antennas when each remote unit is equipped
with an omnidirectional antenna. While the base station can
adaptively steer beams to remote users using $M$ RF chains, we
investigate the Butler matrix, a simple FBN at the RF stage
producing orthogonal beams and requiring only one RF chain for the
best beam to be selected for transmission \cite{but61}. The choice
of the best beam can be achieved with partial channel state
information (CSI) at the base station. The remote feeds back the
index of the best beam to the base station for the forward link.

Although beam selection has been known to have no advantage over
antenna selection in ideal Rayleigh fading channels, it has been
established (using analysis and simulations) that beam selection can
outperform antenna selection in correlated Rayleigh fading channels
with limited angle spread \cite{choi06}. For the case of Rician
fading channels, there exist only limited analytical results of two
very special cases of Rayleigh fading channels and deterministic
channels except our own work in \cite{bai08} while simulations and
measurements have shown that beam selection using the Butler FBN
outperforms antenna selection \cite{grau06}.

Motivated by this, we have analyzed the performance of beam
selection using the Butler FBN for Rician fading channels with
arbitrary $K$-factors and derived the exact distribution of the beam
selection gain as a function of the azimuthal location of the remote
user in our previous work \cite{bai08}, where some key properties of
the noncentral chi-square distribution and the following properties
of the beam selection gain have been presented without any proofs.
Using these properties, we have compared the beam selection gain
with the antenna selection gain for Rician fading channels and
analytically proved that beam selection outperforms antenna
selection.

In this paper, we provide the proofs omitted in \cite{bai08}, which
verify our claim that beam selection is superior to antenna
selection regardless of user location in Rician channels with any
$K$-factors. Moreover, we find asymptotically tight stochastic
bounds of the beam selection gain yielding approximate outage and
the approximate expression for average performance. This
approximation technique can be applied for most of average
performance measures as shown for the expected selection gain and
the ergodic capacity. Using these results, we obtain orders of
growth of the expected selection gain and the ergodic capacity for
beam selection, proved to be higher than those for antenna
selection.

The remainder of this paper is organized as follows: In Section
\ref{sec:model}, we present our system model when the Butler FBN is
used in the base station. In Section \ref{sec:gain}, we analyze the
beam selection gain using a statistical approach. In Section
\ref{sec:gaincom}, we compare the gain of beam selection with that
of antenna selection, and prove that beam selection outperforms
antenna selection under any Rician channel transmission model. In
Section \ref{sec:asymptotic}, we find stochastic bounds of the beam
selection gain and approximate closed form expressions of
performance measures. Finally, we provide our conclusions in Section
\ref{sec:conclusion}.

\section{The System Model}
\label{sec:model}

\begin{figure*}[tb]
\begin{center}
\epsfig{file=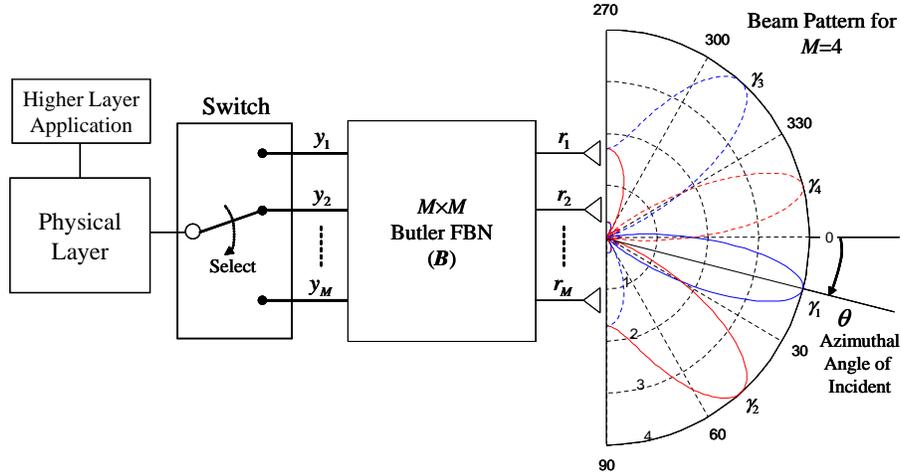,width=120mm}
\caption{Beam selection system using the Butler FBN with $M$ linear
equally spaced array antennas and beam pattern for $M=4$ and $d =
\lambda_c/2$.} \label{fig:fbn}
\end{center}
\end{figure*}

We consider a base station endowed with $M \geq 2$ antennas (as
depicted in Fig.~\ref{fig:fbn}) and remote units each endowed with
one antenna.
For the $m$-th port of the Butler matrix ($m \in \{1,...,M\}$), the
SNR equals to $\rho \cdot {\it\Gamma}_m$ regardless of the direction
of the communication link \cite{bai08}, where $\rho$ is the average
SNR per port and ${\it\Gamma}_m$ denotes the gain of selecting the
$m$-th port. This gain is given by
\begin{equation}
{\it\Gamma}_m = \left|\vec{b}_m^T\vec{h}\right|^2,
\label{eq:SNRgain}
\end{equation}
where the $M \times 1$ complex vector $\vec{h} = [h_1, ..., h_M]^T$
represents the flat fading channel gains for corresponding antennas
normalized such that $\mathbb{E}[|h_i|^2]=1$ for $i=1,2, \cdots, M$,
and the $1 \times M$ complex vector $\vec{b}_m^T$ is the $m$-th row
of the $M \times M$ Butler matrix given by
\begin{equation}
\vec{b}_m^T =
\frac{1}{\sqrt{M}}\left[e^{j\frac{2\pi}{M}\left(m-\frac{1}{2}\right)},
e^{j\frac{2\pi}{M}\left(m-\frac{1}{2}\right)2}, ...,
e^{j\frac{2\pi}{M}\left(m-\frac{1}{2}\right)M}\right].
\end{equation}

Then, the base station chooses the port with the highest SNR. To
select the best beam for the forward link, the remote user only
needs to feed back the index of the best beam to the base station
(even when the channel is not reciprocal) and this is the only
difference between reverse and forward link beam selection. From
this point on, we will not distinguish reverse and forward link beam
selection in this paper as they are analytically identical. The SNR
is then given by $\rho \cdot {\it\Gamma}_{(M)}$, where the notation
$z_{(m)}$ is used to denote the $m$-th smallest value from any set
of finite samples $\{z_1,...,z_M\}$, and thus ${\it\Gamma}_{(M)} =
\max_{m\in\{1,...,M\}} {\it\Gamma}_m$.

%

We define the \textit{beam selection gain} as the ratio of the SNR
of beam selection with a FBN to the average SNR of random antenna
switching without a FBN, which is given by ${\it\Gamma}_{(M)}$.

\section{Beam Selection Gains in Fading Channels}
\label{sec:gain}

It has been shown that beam selection outperforms antenna selection
in ideal line-of-sight (LOS) channels, while beam selection performs
as good as antenna selection in ideal non-line-of-sight (NLOS)
channels \cite{grau06}. We are interested in investigating the
performance of beam selection under Rician channel models. This is
the most frequently used realistic channel model in wireless
communications. Under the Rician channel model, the normalized
channel vector $\vec{h}$ can be modeled as multipath signals
\begin{equation}
\vec{h} = \sqrt{\frac{K}{K+1}}\vec{h}_L +
\sqrt{\frac{1}{K+1}}\vec{h}_N. \label{eq:rician}
\end{equation}
The entries of complex vector $\vec{h}_L$ (which represents the
normalized LOS component) are modeled to have unit power and fixed
phase. The entries of the complex vector $\vec{h}_N$ (which
represents the normalized NLOS component) are modeled by i.i.d.
independent zero-mean circularly symmetric complex Gaussian random
variables with unit variance. The parameter $K$ is referred to as
the Rician $K$-factor, which represents the ratio of the LOS signal
power to the NLOS signal power. The special cases of $K=\infty$ and
$K = 0$ represent ideal LOS (deterministic) and ideal NLOS (Rayleigh
fading) channels, respectively.

\subsection{Deterministic Components}
\label{subsec:los}

Consider the LOS component $\vec{h}_L$. Let $\theta$ denote the
azimuthal angle of incident between a LOS signal and the line
perpendicular to the linear equally spaced array antennas assuming
two-dimensional geometry (horizontal plane) as shown in
Fig.~\ref{fig:fbn}. Furthermore, assume that the distance between
the base station and the mobile user is much larger than array
antenna separation. Then for both reverse and forward link beam
selection, $\vec{h}_L$ is given by
\begin{eqnarray}
\vec{h}_L = \exp(j\psi) \left[1,
\exp\left(-j2\pi\frac{d}{\lambda_c}\sin\theta\right),...,\right. \nonumber\\
\left.
\exp\left(-j2\pi(M-1)\frac{d}{\lambda_c}\sin\theta\right)\right]^T,
\end{eqnarray}
where $\psi$ is an arbitrary phase shift of the signal from/to the
first array antenna, $d$ is the distance between adjacent array
antennas, and $\lambda_c$ is the carrier wavelength.

Let the SNR gain of the $m$-th beam in ideal LOS channels
($K=\infty$) be denoted by
\begin{eqnarray}
\gamma_m &\triangleq& \left|\vec{b}_m^T\vec{h}_L\right|^2 \nonumber \\
&=& \left\{ \begin{array}{l@{\quad}l} M, & \mbox{if } \phi_m = 2\pi n, \;\; n \in \mathbb{Z},\\
\frac{1}{M} \; \frac{\sin^2\left(M\phi_m
/2\right)}{\sin^2\left(\phi_m /2\right)}, & \mbox{otherwise},
\end{array} \right.
\end{eqnarray}
where
\begin{equation}
\phi_m \triangleq 2\pi \left[ \frac{1}{M}\left(m-\frac{1}{2}\right)
- \frac{d}{\lambda_c} \sin \theta \right]. \label{eq:phimdef}
\end{equation}
Since $\vec{h}_L$ is a function of $\theta$, $\gamma_m$ is also a
function of $\theta$ and let us call a set of $M$ functions
$\{\gamma_m|m=1,...,M\}$ a \textit{beam pattern}, which has the
following properties:
\begin{equation}
\sum_{m=1}^{M} \gamma_m = M, \quad 0 \leq \gamma_m \leq M;
\label{eq:gamma_sum}
\end{equation}
\begin{eqnarray}
\gamma_m = M &\mbox{if and only if}& \phi_m = 2\pi \frac{n}{M}, \;
\frac{n}{M} \in \mathbb{Z}; \label{eq:beam_dir}\\
\gamma_m = 0 &\mbox{if and only if}& \phi_m = 2\pi \frac{n}{M}, \;
\frac{n}{M} \notin \mathbb{Z};
\end{eqnarray}
where 
the azimuthal angle satisfying (\ref{eq:beam_dir}) is the
\textit{beam direction}. Let us define a lobe of a beam as a
\textit{main lobe} if the beam direction is inside that lobe.
We assume
\begin{equation}
\frac{M-1}{2M} < \frac{d}{\lambda_c}, \label{eq:min_sep}
\end{equation}
for all $M$ beams to have at least one main lobe. We examine the
beam pattern only from $\theta = 0$ to the first beam direction
given by
\begin{equation}
\theta = \nu \triangleq \arcsin
\left(\frac{1}{2M}\frac{\lambda_c}{d}\right)
\end{equation}
as discussed in \cite{bai08}.
%

\subsection{Probabilistic Analysis}
\label{subsec:stat}

Now, let us consider the statistical channel model including NLOS
components. 
The cumulative distribution function (cdf) of ${\it\Gamma}_m$ is
given by \cite{bai08}
\begin{eqnarray}
F_m (x) &\triangleq& \Pr \{ {\it\Gamma}_m \leq x \} \nonumber
\\
&=& \left. F_{\chi'^2}
(2(K+1)x|n,\delta)\right|_{n=2,\;\delta=2K\gamma_m} \nonumber \\
& 
=& 
\left.
\mathbb{E} \left[F_{\chi^2}
(2(K+1)x|n+2P_{\delta/2})\right]\right|_{n=2,\;\delta=2K\gamma_m},
\label{eq:SNRdist}
\end{eqnarray}
where $F_{\chi'^2} (x|n,\delta)$ is the noncentral chi-square cdf
with $n$ degrees of freedom and the noncentrality parameter
$\delta$, $P_{\delta/2}$ is a Poisson random variable with mean
$\delta/2$, and $F_{\chi^2} (x|q)$ is the chi-square cdf with $q$
degrees of freedom, given by
\begin{equation}
F_{\chi^2} (x|q) = 1-e^{-x/2}\sum_{k=0}^{q/2-1} \frac{(x/2)^k}{k!} =
e^{-x/2}\sum_{k=q/2}^{\infty} \frac{(x/2)^k}{k!} \label{eq:chisq}
\end{equation}
if $q$ is an even number as in (\ref{eq:SNRdist}) where
$q=n+2P_{\delta/2}|_{n=2}$. Note that 
given $K$, evaluating $\gamma_m$ is enough to know the distribution
of the SNR gain ${\it\Gamma}_m$. 
The beam selection gain ${\it\Gamma}_{(M)}$ is given by
\begin{eqnarray}
F_{(M)} (x) &\triangleq& \Pr \{ {\it\Gamma}_{(M)} \leq x \} 
= \prod_{m=1}^{M} F_m (x),
\end{eqnarray}
and thus for $x > 0$,
\begin{equation}
\log F_{(M)} (x) = \sum_{m=1}^{M} \log F_m (x).
\end{equation}

We have the following useful key theorem on the noncentral
chi-square distribution, whose proof can be found in the Appendix. 

\begin{theorem}
\label{th:concavity} 
The logarithm of the noncentral chi-square cdf with two degrees of
freedom
\begin{equation}
\log F_{\chi'^2} (x|2,\delta) \label{eq:logncd}
\end{equation}
is a strictly decreasing and strictly concave function of the
noncentrality parameter $\delta \geq 0$ for any given $x > 0$
assuming that the base of logarithm is greater than one. 
\thmend
\end{theorem}

%

Now, we are ready to show the following theorem, where stochastic
order relations are introduced in \cite[Ch.~9]{ross96}.

\begin{theorem}
\label{th:stochorder} 
For any given $x > 0$, $F_{(M)} (x)$, the cdf of the beam selection
gain ${\it\Gamma}_{(M)}$, is a strictly decreasing function of
$\theta$ from zero to the first beam direction $\nu = \arcsin
\left(\frac{1}{2M}\frac{\lambda_c}{d}\right)$. Therefore, in this
interval, ${\it\Gamma}_{(M)}$ is stochastically increasing,
stochastically smallest at $\theta = 0$, and stochastically largest
at $\theta = \nu$. 
\thmend
\end{theorem}

\begin{proof}
This proof is given in the Appendix.
\end{proof}


The corollary below follows naturally from Theorem
\ref{th:stochorder}.

\begin{corollary}
\label{cor:bounds} 
For $\theta \in [-\pi/2, \pi/2]$ and any integer $|m| \leq
\frac{M}{\lambda_c / d}$, ${\it\Gamma}_{(M)}$ is stochastically
increasing as $\theta$ increases if
\begin{eqnarray}
\theta &\in& \left[\arcsin
\left(\frac{m}{M}\frac{\lambda_c}{d}\right),\right.\nonumber\\
& & \arcsin
\left.\left(\min\left\{\frac{m+1/2}{M}\frac{\lambda_c}{d},1
\right\}\right)\right],
\end{eqnarray}
and stochastically decreasing as $\theta$ increases if
\begin{eqnarray}
\theta &\in& \left[\arcsin
\left(\max\left\{\frac{m-1/2}{M}\frac{\lambda_c}{d},
-1\right\}\right),\right.\nonumber\\
& & \hspace{1cm}
\left.\arcsin\left(\frac{m}{M}\frac{\lambda_c}{d}\right)\right].
\end{eqnarray}
It is exactly opposite for the other half of the horizontal plane,
$\theta \in [\pi/2, 3\pi/2]$. Therefore, ${\it\Gamma}_{(M)}$ with
$\theta = 0$ and $\theta = \nu$ are achievable stochastic lower and
upper bounds, respectively for ${\it\Gamma}_{(M)}$ with an arbitrary
$\theta$. 
\thmend
\end{corollary}

Corollary \ref{cor:bounds} tells us that the expected performance
measures over ${\it\Gamma}_{(M)}$ with $\theta = 0$ and $\theta =
\nu$ can serve as lower and upper bounds, respectively, for the
averages of any performance measures which are increasing functions
of SNR, e.g., the channel capacity. They can also serve as upper and
lower bounds, respectively, for the averages of any performance
measures which are decreasing functions of SNR, e.g., the bit error
rate (BER), applying the result in \cite[pp.~405--406]{ross96}.

\section{Beam Selection Versus Antenna Selection} \label{sec:gaincom}


Let us consider the antenna selection gain under the same scenario
used for beam
selection case  
except the fact that the Butler FBN will not be deployed for antenna
selection. When the $m$-th antenna is selected among $M$ antennas in
the base station, the SNR is given by $\rho \cdot H_m$, where $H_m
\triangleq |h_m|^2$. 
Assuming that the antenna with the highest SNR is always selected,
the antenna selection gain is defined as the ratio of the SNR of
antenna selection to the average SNR of random antenna switching,
which can be expressed by $H_{(M)}$. For any $m$, the cdf of $H_m$
becomes
\begin{equation}
G (x) \triangleq \Pr \{ H_m \leq x \} = F_{\chi'^2} (2(K+1)x|2,2K).
\end{equation}
Therefore, the cdf of $H_{(M)}$ is given by
\begin{equation}
G_{(M)} (x) \triangleq \Pr \{ H_{(M)} \leq x \} = G^M (x).
\end{equation}
With the proofs of previous theorems, we can confirm that the
following lemma holds.

\begin{lemma}
\label{lem:beamvsant} 
For the same Rician $K$-factor, beam selection always outperforms
antenna selection, i.e., the beam selection gain ${\it\Gamma}_{(M)}$
is stochastically larger than the antenna selection gain $H_{(M)}$.
\thmend
\end{lemma}

\begin{proof}
Applying the concavity result in Theorem \ref{th:concavity} and
Jensen's inequality gives us
\begin{eqnarray}
\lefteqn{\log G_{(M)} (x) = M \log F_{\chi'^2} (2(K+1)x|2,2K)} \nonumber \\
&\geq& \sum_{m=1}^{M} \log F_{\chi'^2} (2(K+1)x|2,2K\gamma_m)
\nonumber \\
&=& \log F_{(M)} (x),
\end{eqnarray}
for any given $x > 0$.
\end{proof}


\section{Asymptotic Selection Gains}
\label{sec:asymptotic}

It has been shown that the beam selection gain is stochastically
upper and lower bounded by ${\it\Gamma}_{(M)}$ with $\theta$ of zero
and the first beam direction $\nu = \arcsin
\left(\frac{1}{2M}\frac{\lambda_c}{d}\right)$, respectively. Our
interest in this section is to see how these two extremes change as
the number of antennas $M$ increases and then obtain the asymptotic
selection gain for an arbitrary location of the remote user.
Furthermore, these analytical results can be applied to study the
outage and the ergodic capacity of beam selection systems. For this
purpose, consider the SNR gain ${\it\Gamma}_m(\theta)$ and its cdf
$F_m(\cdot|\theta)$ as functions of the azimuthal angle $\theta$.

\subsection{Bounds and Approximations} \label{subsec:bounds}

First, we can obtain the stochastic lower bound for the beam
selection gain of the user at the beam direction
${\it\Gamma}_{(M)}(\nu)$ given by
\begin{eqnarray}
\lefteqn{F_{(M)}(x|\nu) 
= \prod_{m=1}^{M} F_{\chi'^2} (2(K+1)x|2,2K\gamma_m(\nu))}
\nonumber \\
&=& F_{\chi'^2} (2(K+1)x|2,2KM) \cdot F^{M-1}_{\chi^2} (2(K+1)x|2) \nonumber \\
&=& Q_{M} (x) W^{M-1} (x) \nonumber \\
&\leq& Q_M (x), \label{eq:maxbound}
\end{eqnarray}
where $Q$ and $W$ are defined by
\begin{eqnarray}
Q_{\gamma} (x) &\triangleq& F_{\chi'^2}
(2(K+1)x|2,2K\gamma), \\
W (x) &\triangleq& F_{\chi^2} (2(K+1)x|2).
\end{eqnarray}

\begin{figure}[tb]
\begin{center}
\epsfig{file=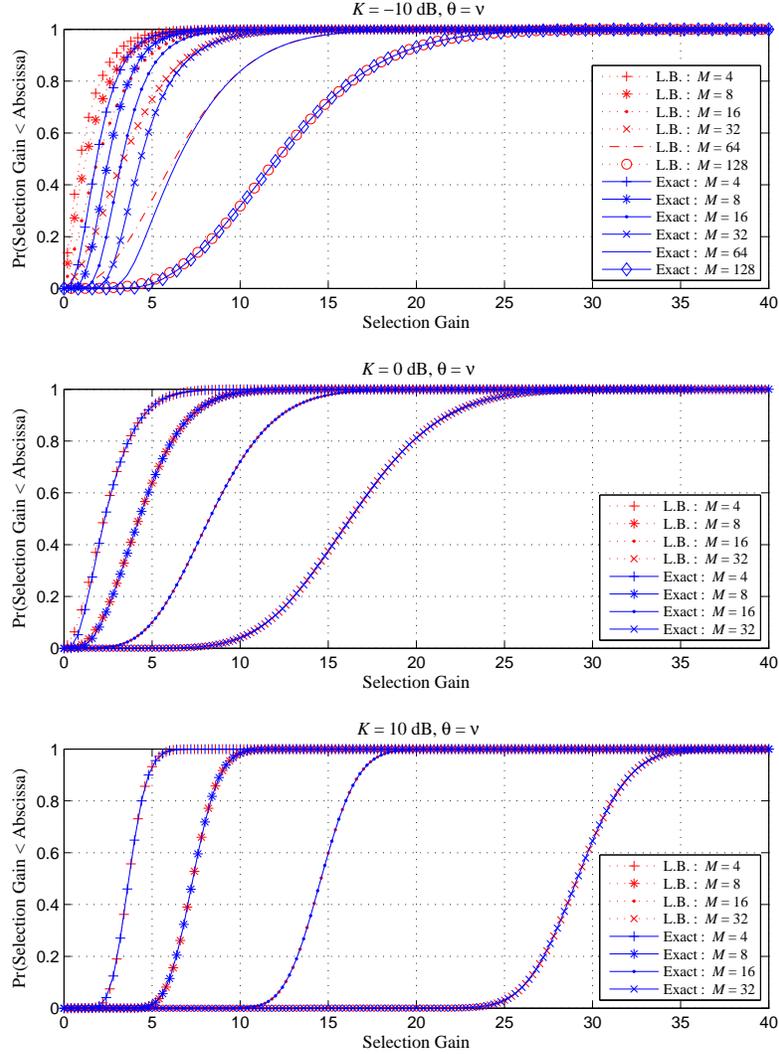,width=88.9mm,trim=1.4cm
0cm 1.4cm 0cm}
\caption{Distributions of the beam selection gain
${\it\Gamma}_{(M)}(\nu)$ and its stochastic lower bound for $K =
-10,\,0,\,10$\,dB, where $d = \lambda_c/2$ is assumed.}
\label{fig:beammaxbound}
\end{center}
\end{figure}
Fig.~\ref{fig:beammaxbound} shows $F_{(M)}(x|\nu)$ and its
stochastic lower bound $Q_M$. It can be seen that the lower bound
$Q_M$ approaches to the cdf $F_{(M)}(x|\nu)$ as $M$ increases, which
will be proved.


Now, consider the beam selection gain of the user exactly between
beams ${\it\Gamma}_{(M)}(0)$ and its cdf given by
\begin{equation}
F_{(M)}(x|0) = \prod_{m=1}^{M} F_{\chi'^2}
(2(K+1)x|2,2K\gamma_m(0)). \label{eq:lowerdef}
\end{equation}
Let us choose a vector $\vec{u} = [u_{(1)},...,u_{(M)}]$ which
majorizes the beam pattern $\{\gamma_m(0)|m=1,...,M\}$ as
\begin{eqnarray}
& & \gamma_1(0) = \gamma_M(0) = u_{(M)} = u_{(M-1)} = \frac{1}{M
\sin^2
(\pi/2M)} \triangleq a_M \nonumber \\
& &> u_{(M-2)} = u_{(M-3)} = \frac{M}{2} - \frac{1}{M \sin^2
(\pi/2M)}
\triangleq b_M > \gamma_2(0) = \gamma_{M-1}(0) \nonumber \\
& &> \gamma_3(0),...,\gamma_{M-2}(0) > u_{(M-4)} = ... = u_{(1)} =
0,
\end{eqnarray}
where majorization is introduced in \cite[p.~45]{hardy52}.

{\it Notation:} For any two real-valued sequences $c_M$ and $d_M$,
we define
\[ \begin{array}{lll}
c_M \approx d_M & \mbox{if and only if} & \lim_{M \rightarrow
\infty} |c_M - d_M| = 0; \\
c_M \sim d_M & \mbox{if and only if} & \lim_{M \rightarrow \infty}
c_M / d_M = 1; \\
c_M = \mathnormal{\Theta}(d_M) & \mbox{if and only if} & 0 < \lim_{M
\rightarrow \infty} c_M / d_M < \infty.
\end{array} \]
\thmend

Using this notation, we can see
\begin{eqnarray}
a_M &\sim& \frac{4}{\pi^2} M = (0.4053...) \times M, \label{eq:aMgrowth} \\
b_M &\sim& \left(\frac{1}{2} - \frac{4}{\pi^2}\right) M =
(0.0947...) \times M. \label{eq:bMgrowth}
\end{eqnarray}
Applying Hardy-Littlewood-P\'{o}lya¡¯s theorem in
\cite[pp.~88--91]{hardy52} and the strict concavity of
(\ref{eq:logncd}) to (\ref{eq:lowerdef}) yields the stochastic upper
bound
\begin{eqnarray}
\lefteqn{F_{(M)}(x|0) \geq \prod_{m=1}^{M} F_{\chi'^2}
(2(K+1)x|2,2Ku_{(m)})} \nonumber \\
&=& F_{\chi'^2}^2 (2(K+1)x|2,2K a_M) \cdot F_{\chi'^2}^2
(2(K+1)x|2,2K b_M) \cdot F_{\chi^2}^{M-4} (2(K+1)x|2) \nonumber \\
&=& Q_{a_M}^2 (x) \cdot Q_{b_M}^2 (x) \cdot W^{M-4} (x).
\end{eqnarray}
Thus, we have the stochastic lower and upper bound for
$F_{(M)}(x|0)$ given by
\begin{equation}
Q^2_{a_M} (x) \geq F_{(M)}(x|0) \geq Q^2_{a_M} (x) \cdot Q^2_{b_M}
(x) \cdot W^{M-4} (x). \label{eq:minbound}
\end{equation}

\begin{figure}[tb]
\begin{center}
\epsfig{file=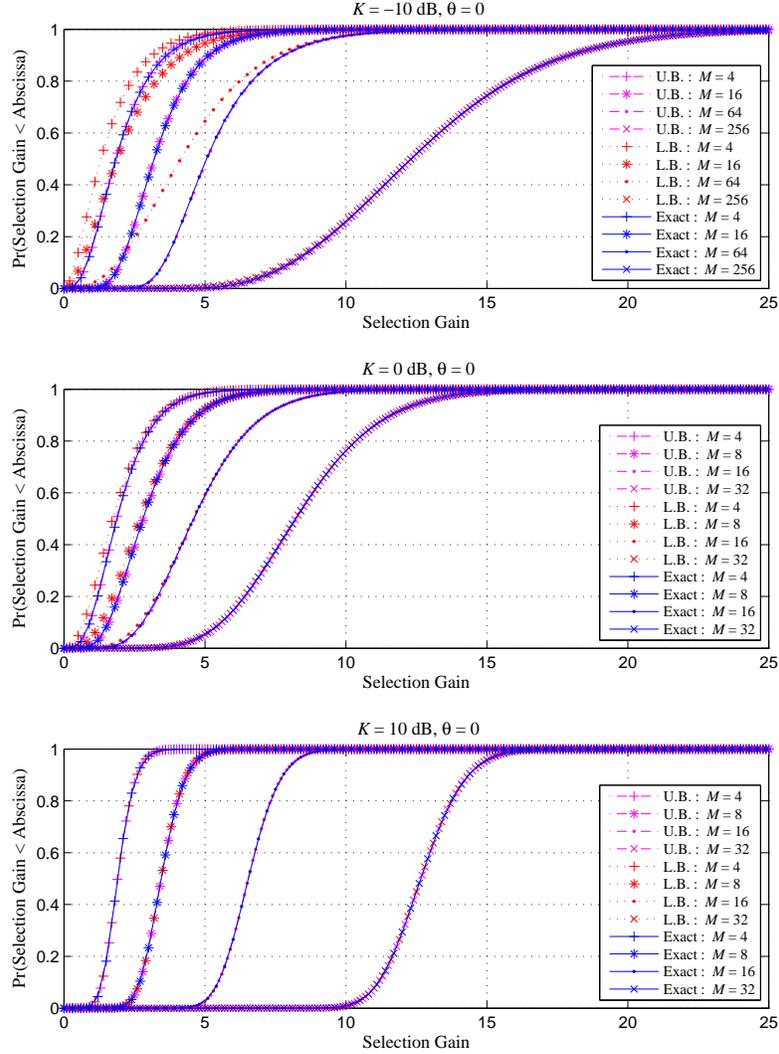,width=88.9mm,trim=1.4cm
0cm 1.4cm 0cm}
\caption{Distributions of the beam selection gain
${\it\Gamma}_{(M)}(0)$ and its stochastic lower and upper bounds for
$K = -10,\,0,\,10$\,dB, where $d = \lambda_c/2$ is assumed.}
\label{fig:beamminbound}
\end{center}
\end{figure}
Fig.~\ref{fig:beamminbound} shows $F_{(M)}(x|0)$ and its stochastic
lower bound $Q^2_{a_M}$ and upper bound  $Q^2_{a_M} Q^2_{b_M}
W^{M-4}$. We also observe that as the lower and upper bounds are
merged into each other, so does $F_{(M)}(x|0)$ as $M$ increases.

The following theorem verifies that the stochastic lower bounds in
(\ref{eq:maxbound}) and (\ref{eq:minbound}) are indeed
asymptotically tight.

\begin{theorem}
\label{th:tight} 
For $K > 0$ and $p \in [0,1)$,
\begin{equation}
F_{(M)}^{-1}(p|\nu) \approx  Q_M^{-1} (p) \label{eq:uppertight}
\end{equation}
and
\begin{equation}
F_{(M)}^{-1}(p|0) \approx Q_{a_M}^{-1} (\sqrt{p})
\label{eq:lowertight}
\end{equation}
as $M$ increases.
\thmend
\end{theorem}

\begin{proof}
This proof is given in the Appendix.
\end{proof}

We also have the following theorem useful for average performance
evaluation, whose proof can be found in the Appendix.

\begin{theorem}
\label{th:meanconverg} 
Let $h$ be any differentiable function defined on $[0,\infty)$ such
that $h'$ is bounded. If $h$ is integrable with respect to $Q_M$,
then
\begin{equation}
\int_0^{\infty} h(x) d F_{(M)}(x|\nu) \approx  \int_0^{\infty} h(x)
d Q_M (x) \label{eq:firstmeanconv}
\end{equation}
as $M$ increases. If $h$ is integrable with respect to $Q_{a_M}^2$,
then
\begin{equation}
\int_0^{\infty} h(x) d F_{(M)}(x|0) \approx  \int_0^{\infty} h(x) d
Q_{a_M}^2 (x) \label{eq:secondmeanconv}
\end{equation}
as $M$ increases.
\thmend
\end{theorem}

Theorems \ref{th:tight} and \ref{th:meanconverg} in this subsection
demonstrate that for large $M$, the distributions of the beam
selection gain of the user at the beam direction
${\it\Gamma}_{(M)}(\nu)$ and the beam selection gain of the user
exactly between beams ${\it\Gamma}_{(M)}(0)$ can be well
approximated by $Q_M (x)$ and $Q_{a_M}^2 (x)$, respectively, which
are the noncentral chi-square distribution and its square. These are
useful as their closed-form expressions are complicated and thus not
insightful.

\subsection{Performance Analysis} \label{subsec:bounds}

It can be seen that outage probabilities with $\theta = 0$ and
$\theta = \nu$ for a given rate $C_0$ can be approximated by
\begin{eqnarray}
P_{out}(C_0) &\triangleq& \Pr
\left\{\log_2\left(1+\rho\Gamma_{(M)}(\theta)\right) \leq C_0 \right\} \nonumber \\
&\approx& \left\{ \begin{array}{ll}
Q_M\left(\frac{2^{C_0}-1}{\rho}\right),
& \mbox{if } \theta = \nu, \\
Q^2_{a_M}\left(\frac{2^{C_0}-1}{\rho}\right), & \mbox{if } \theta =
0,\end{array} \right.
\end{eqnarray}
for large $M$. Furthermore, Theorem \ref{th:tight} can be used to
approximate outage capacities with $\theta = 0$ and $\theta = \nu$
as
\begin{eqnarray}
C_{\rm{out}}(P_0) &\triangleq & P_{\rm{out}}^{-1} (P_0)
= \log_2 \left[1+\rho F^{-1}_{(M)}(P_0 | \theta)\right] \nonumber \\ 
&\approx& \left\{ \begin{array}{ll} \log_2 \left[1+\rho
Q^{-1}_M(P_0)\right], & \mbox{if } \theta = \nu, \\
\log_2 \left[1+\rho Q^{-1}_{a_M}(\sqrt{P_0})\right], & \mbox{if }
\theta = 0.\end{array} \right.
\end{eqnarray}
for large $M$.

Let us apply Theorem \ref{th:meanconverg} to the mean selection gain
$\mathbb{E}\left[{\it\Gamma}_{(M)}\right]$ by taking $h(x) = x$. The
expected beam selection gain for $\theta = \nu$ is given by
\begin{equation}
\mathbb{E}\left[{\it\Gamma}_{(M)}(\nu)\right] \approx
\int_0^{\infty} x d Q_M (x) = \frac{KM+1}{K+1} 
= \mathnormal{\Theta}(M). \label{eq:meanQM}
\end{equation}
The expected beam selection gain for $\theta = 0$ is given by
\begin{equation}
\mathbb{E}\left[{\it\Gamma}_{(M)}(0)\right] \approx \int_0^{\infty}
x d Q_{a_M}^2 (x), \label{eq:lowermeangain}
\end{equation}
as $M$ increases. Although it seems difficult to solve the
integration in (\ref{eq:lowermeangain}), we can obtain upper and
lower bounds using an inequality in \cite[p.~62]{david03} because
$Q_{a_M}^2$ is the cdf of the maximum of two samples from $Q_{a_M}$,
whose mean and variance are $(Ka_M+1)/(K+1)$ and
$(2Ka_M+1)/(K+1)^2$, respectively. These bounds are given by
\begin{equation}
\frac{Ka_M+1}{K+1} \leq \int_0^{\infty} x d Q_{a_M}^2 (x) \leq
\frac{Ka_M+1}{K+1} + \frac{1}{\sqrt{3}} \frac{\sqrt{2Ka_M+1}}{K+1},
\label{eq:meanQsqaM}
\end{equation}
which yields
\begin{equation}
\mathbb{E}\left[{\it\Gamma}_{(M)}(0)\right] \approx \int_0^{\infty}
x d Q_{a_M}^2 (x) \sim \frac{K a_M +1}{K+1}  
= \mathnormal{\Theta}(M)
\end{equation}
Hence, $\mathbb{E}\left[{\it\Gamma}_{(M)}\right] =
\mathnormal{\Theta}(M)$ regardless of user location, which is faster
than $\mathnormal{\Theta}(\log M)$ for antenna selection
\cite{bai07}.

\begin{lemma}
\label{lm:capconverg} 
Let $\rho > 0$ denote SNR. As $M$ increases, the ergodic capacity of
the user at the beam direction ($\theta = \nu$) is given by
\begin{equation}
\mathbb{E} \left[ \log_2\left(1+\rho {\it\Gamma}_{(M)}(\nu)\right)
\right] \approx \log_2\left(1+\rho\frac{KM+1}{K+1} \right),
\label{eq:maxergcapconv}
\end{equation}
and the ergodic capacity of the user exactly between beams ($\theta
= 0$) is given by
\begin{equation}
\mathbb{E} \left[ \log_2\left(1+\rho {\it\Gamma}_{(M)}(0)\right)
\right] \approx \log_2\left(1+\rho\frac{Ka_M+1}{K+1} \right).
\label{eq:minergcapconv}
\end{equation}
\thmend
\end{lemma}

\begin{proof}
This proof is given in the Appendix.
\end{proof}

This lemma also yields the order of growth of the ergodic capacity
$\mathbb{E} \left[ \log_2\left(1+\rho {\it\Gamma}_{(M)}\right)
\right] \approx \mathnormal{\Theta}(\log(M))$ regardless of user
location, which is faster than $\mathnormal{\Theta}(\log(\log(M)))$
for antenna selection \cite{bai07}.
\begin{figure}[tb]
\begin{center}
\epsfig{file=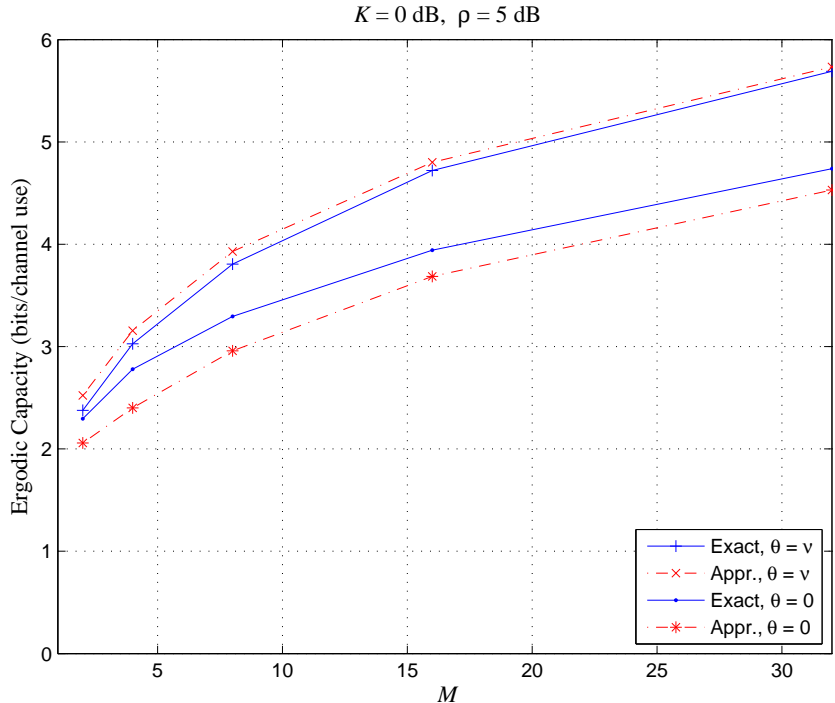,width=88.9mm,trim=1.4cm 0cm
1.4cm 0cm}
\caption{Ergodic capacity versus $M$ for $K = 0$\,dB at $\rho =
5$\,dB, where $d = \lambda_c/2$ is assumed.} \label{fig:ergcapacity}
\end{center}
\end{figure}
Fig.~\ref{fig:ergcapacity} shows the ergodic capacity and its
approximations in (\ref{eq:maxergcapconv}) and
(\ref{eq:minergcapconv}) for SNR $\rho = 5$\,dB. We see that the
approximations approach the numerically integrated exact values as
$M$ increases.

\section{Conclusion}
\label{sec:conclusion}

We considered beam selection using the Butler FBN at the base
station with multiple linear equally spaced omnidirectional array
antennas. Completing the analysis of the beam selection gain, we
provided the proofs of the key properties verifying that beam
selection is superior to antenna selection in Rician channels with
any $K$-factors. We also found asymptotically tight stochastic
bounds of the beam selection gain and approximate closed form
expressions of the expected selection gain and the ergodic capacity.
Using these results, it was shown that beam selection has higher
order of growth of the ergodic capacity than antenna selection.
Graphical results were provided demonstrating the underlying gains
and supporting our approximations.

\appendix

\myproof{Proof of Theorem \ref{th:concavity}:} 
Without loss of generality, assume the natural logarithm. For any
given $x
> 0$, (\ref{eq:logncd}) can be expressed as
\begin{eqnarray}
\log F_{\chi'^2} (x|2,\delta) &=& \log \left[\sum_{i=0}^{\infty}
\frac{e^{-\delta/2}(\delta/2)^i}{i!} \alpha_i \right] \nonumber
\\
&=& -\frac{\delta}{2} + \log \left[\sum_{i=0}^{\infty}
\frac{(\delta/2)^i}{i!} \alpha_i\right], \label{eq:logncdetail}
\end{eqnarray}
where $\alpha_i$ is defined as
\begin{equation}
\alpha_i \triangleq F_{\chi^2} (x|2+2i) = e^{-x/2}
\sum_{k=i+1}^{\infty} \frac{(x/2)^k}{k!} \label{eq:alphadef}
\end{equation}
from (\ref{eq:chisq}). Differentiating (\ref{eq:logncdetail}) gives
us
\begin{eqnarray}
\frac{\partial}{\partial\delta}\log F_{\chi'^2} (x|2,\delta) &=&
-\frac{1}{2}+\frac{1}{2} \cdot \frac{\sum_{i=0}^{\infty}
\frac{(\delta/2)^i}{i!} \alpha_{i+1}}{\sum_{i=0}^{\infty}
\frac{(\delta/2)^i}{i!}
\alpha_i} \nonumber \\
&\hspace{-1.8cm}=& \hspace{-1cm} \frac{\sum_{i=0}^{\infty}
\frac{(\delta/2)^i}{i!} (\alpha_{i+1}-\alpha_{i})}{2
\sum_{i=0}^{\infty} \frac{(\delta/2)^i}{i!} \alpha_i}< 0
\end{eqnarray}
for $\delta > 0$ because $\alpha_{i+1} < \alpha_i$ from
(\ref{eq:alphadef}), and thus (\ref{eq:logncd}) is a strictly
decreasing function of $\delta \geq 0$.

Now, prove that (\ref{eq:logncd}) is a strictly concave function of
$\delta \geq 0$. The second derivative of (\ref{eq:logncdetail}) is
given by
\begin{eqnarray}
\lefteqn{\frac{\partial^2}{\partial\delta^2}\log F_{\chi'^2}
(x|2,\delta)}
\nonumber \\
&\hspace{-1.8cm}=& \hspace{-1cm} \frac{\left(\sum_{i=0}^{\infty}
\frac{(\delta/2)^i}{i!} \alpha_i \right)\left(\sum_{i=0}^{\infty}
\frac{(\delta/2)^i}{i!} \alpha_{i+2} \right) -
\left(\sum_{i=0}^{\infty} \frac{(\delta/2)^i}{i!} \alpha_{i+1}
\right)^2}{4\left(\sum_{i=0}^{\infty} \frac{(\delta/2)^i}{i!}
\alpha_i \right)^2}, \label{eq:2nddiff}
\end{eqnarray}
the $i$-th order term of whose numerator can be simplified as
\begin{equation}
\frac{(\delta/2)^i}{i!} (\alpha_0 \alpha_{i+2} - \alpha_1
\alpha_{i+1}). \label{eq:alphacond}
\end{equation}
Let us show that (\ref{eq:2nddiff}) is negative by proving that
(\ref{eq:alphacond}) is negative for $\delta > 0$. Consider
$\alpha_{i-1} /\alpha_i$, which is an increasing function of $i$
because
\begin{eqnarray}
\frac{\alpha_{i-1}}{\alpha_i} - 1 &=&
\frac{\frac{(x/2)^i}{i!}}{\sum_{k=i+1}^{\infty} \frac{(x/2)^k}{k!}}
\nonumber \\
&=& \frac{1}{\sum_{k=1}^{\infty} \frac{(x/2)^k}{(i+k)!/i!}}
\end{eqnarray}
and $(i+k)!/i!$ increases as $i$ increases for any positive integer
$k$. Therefore,
\begin{equation}
\frac{\alpha_0}{\alpha_1} < \frac{\alpha_1}{\alpha_2} < ... <
\frac{\alpha_{i+1}}{\alpha_{i+2}} < ...,
\end{equation}
which yields the strict concavity of (\ref{eq:logncd}).
\endproof

\myproof{Proof of Theorem \ref{th:stochorder}:} 
Define
\begin{equation}
\beta \triangleq 2\pi \frac{d}{\lambda_c} \sin \theta.
\end{equation}
Under the condition (\ref{eq:min_sep}), $\beta$ is an increasing and
continuous function of $\theta$ and has the range $\left[0,
\frac{\pi}{M}\right]$. Therefore, we only need to show that
$F_{(m)}$ is a strictly decreasing function of $\beta$ in the domain
$\left[0, \frac{\pi}{M}\right]$. From (\ref{eq:phimdef}),
\begin{equation}
\phi_m (\beta) \triangleq \frac{2\pi}{M}\left(m-\frac12\right) -
\beta = \phi_m (0) - \beta,
\end{equation}
and by defining
\begin{equation}
\eta(\phi) = \left\{ \begin{array}{l@{\quad}l} M & \mbox{if } \phi = 2\pi n, \;\; n \in \mathbb{N} \\
\frac{1}{M} \; \frac{\sin^2\left(M\phi/2\right)}{\sin^2\left(
\phi/2\right)} 
& \mbox{otherwise},
\end{array} \right. \label{eq:gamdef}
\end{equation}
we can represent
\begin{equation}
\gamma_m (\beta) = \eta(\phi_m(\beta)).
\end{equation}
Note that $\eta(\phi)$ is a periodic function with period $2\pi$,
symmetric with respect to the axis $\phi = \pi n$, and the value of
$\eta$ at $\phi = 2\pi n$ makes $\eta(\phi)$ a continuous function
of $\phi$.

First, prove that for $\beta \in \left[0, \frac{\pi}{M}\right]$ the
beam pattern $\{\gamma_m\}$ can be sorted in nonincreasing order as
follows:
\begin{equation}
\gamma_1(\beta) \geq \gamma_M(\beta) \geq \gamma_2(\beta) \geq
\gamma_{M-1}(\beta) \geq ... \geq
\gamma_{\left\lfloor\frac{M}{2}\right\rfloor+1}(\beta),
\label{eq:gammaorder}
\end{equation}
where $\lfloor\cdot\rfloor$ is a floor function. It can be easily
shown that
\begin{equation}
\gamma_{M+1-m} (\beta) = 
\eta(\phi_m(-\beta)) = \gamma_m(-\beta). \label{eq:gamma_m_sym}
\end{equation}
We get the following equivalent inequalities of
(\ref{eq:gammaorder})
\begin{equation}
\eta(\phi_1(\beta)) \geq \eta(\phi_1(-\beta)) \geq ... \geq
\eta\left(\phi_{\left\lfloor\frac{M+1}{2}\right\rfloor}\left((-1)^{M-1}\beta\right)\right).
\label{eq:gammaequivorder}
\end{equation}
We can see that
\begin{equation}
\eta(\phi_m(\pm\beta)) = \frac{1}{M} \;
\frac{\sin^2\left(\frac{M}{2}
\phi_m(\pm\beta)\right)}{\sin^2\left(\frac{1}{2}
\phi_m(\pm\beta)\right)} = \frac{1}{M} \;
\frac{\cos^2\left(\frac{M}{2} \beta\right)}{\sin^2\left(\frac{1}{2}
\phi_m(\pm\beta)\right)} \label{eq:gamequalnum}
\end{equation}
and
\begin{equation}
0 \leq \phi_1(\beta) \leq \phi_1(-\beta) \leq ... \leq
\phi_{\left\lfloor\frac{M+1}{2}\right\rfloor}\left((-1)^{M-1}\beta\right)
\leq \pi, \label{eq:phiorder}
\end{equation}
which yields (\ref{eq:gammaequivorder}) because in
(\ref{eq:gamequalnum}), the numerator
$\sin^2\left(\frac{M}{2}\phi\right)$ has the same value at
$\phi=\phi_m(\pm\beta)$ for any fixed $\beta$ and all $m$, and the
denominator $\sin^2\left(\frac{1}{2}\phi\right)$ is increasing
function of $\phi \in \left[0, \pi\right]$. Define the
nondecreasingly sorted vector $\vec{\gamma}$ from $\{\gamma_m\}$
given by
\begin{eqnarray}
{\vec{\gamma}} &\triangleq& [\gamma_{(1)}, \gamma_{(2)}, ..., \gamma_{(M)}] \nonumber \\
&=& \left[\gamma_{\left\lfloor\frac{M}{2}\right\rfloor+1},...,
\gamma_{M-1}, \gamma_2, \gamma_M ,\gamma_1 \right]
\end{eqnarray}
for $\beta \in \left[0, \frac{\pi}{M}\right]$. Let us show that
$\vec{\gamma}(\beta_2)$ strictly majorizes $\vec{\gamma}(\beta_1)$
for $0 \leq \beta_1 < \beta_2 \leq \frac{\pi}{M}$, which means
\begin{equation}
\sum_{i=1}^{M} \gamma_{(i)}(\beta_1) = \sum_{i=1}^{M}
\gamma_{(i)}(\beta_2) \label{eq:v_sum}
\end{equation}
and
\begin{equation}
\sum_{i=1}^{m} \gamma_{(i)}(\beta_1) > \sum_{i=1}^{m}
\gamma_{(i)}(\beta_2) \label{eq:majorize}
\end{equation}
for all $m \in \{1,...,M-1\}$. We already have (\ref{eq:v_sum}) from
(\ref{eq:gamma_sum}), and thus it suffices to prove
(\ref{eq:majorize}). Under the assumption that (\ref{eq:majorize})
is proved, using Hardy-Littlewood-P\'{o}lya¡¯s theorem in
\cite[pp.~88--91]{hardy52} based on the strict concavity of
(\ref{eq:logncd}) proved in Theorem \ref{th:concavity} gives us
\begin{eqnarray}
\lefteqn{\log F_{(M)} (x|K,\beta_1) = \sum_{m=1}^{M} \log
F_{\chi'^2}
(2(K+1)x|2,2K\gamma_m(\beta_1))} \nonumber \\
&>& \sum_{m=1}^{M} \log F_{\chi'^2} (2(K+1)x|2,2K\gamma_m(\beta_2))
= \log F_{(M)} (x|K,\beta_2), \label{eq:noncenineq}
\end{eqnarray}
which basically shows that $F_{(m)}$ is a strictly decreasing
function of $\beta$.

Let us prove that $\gamma_1(\beta)$ and $\gamma_M(\beta)$ are
strictly increasing and strictly decreasing respectively.
For $\phi \neq 2\pi n$, it can be shown that
\begin{equation}
\eta'(\phi) = \frac{1}{M} \; \frac{\sin^2\left(\frac{M}{2}
\phi\right)}{\sin^2\left(\frac{1}{2} \phi\right)}
\left[M\cot\left(\frac{M}{2}\phi\right) -
\cot\left(\frac{1}{2}\phi\right) \right].
\end{equation}
We can show $\eta'(\phi)$ is negative for $0 < \phi <
\frac{2\pi}{M}$ because by the Taylor series expansion,
\begin{eqnarray}
\lefteqn{M\cot\left(\frac{M}{2}\phi\right) -
\cot\left(\frac{1}{2}\phi\right)} \nonumber \\
&=& M\left[\frac{2}{M\phi} - \sum_{i=1}^{\infty}
\frac{2^{2i}|B_{2i}|}{(2i)!}\left(\frac{M}{2}\phi\right)^{2i-1}\right]
- \left[\frac{2}{\phi} - \sum_{i=1}^{\infty}
\frac{2^{2i}|B_{2i}|}{(2i)!}\left(\frac{1}{2}\phi\right)^{2i-1}\right] \nonumber \\
&=& - \sum_{i=1}^{\infty}
\frac{2^{2i}|B_{2i}|}{(2i)!}(M^{2i}-1)\left(\frac{\phi}{2}\right)^{2i-1}
< 0 
\end{eqnarray}
where $B_{i}$ is the $i$-th Bernoulli number. Therefore,
$\eta(\phi)$ is strictly decreasing in $\left[0,
\frac{2\pi}{M}\right]$, and thus $\eta(\phi)$ is strictly increasing
in $\left[2\pi\frac{M-1}{M}, 2\pi\right]$ by the symmetry. Since
\begin{equation}
\gamma_1(\beta) 
= \eta \left(\frac{\pi}{M} - \beta \right)
\end{equation}
and
\begin{equation}
\gamma_M(\beta) 
= \eta \left(2\pi\frac{M-1/2}{M} - \beta \right),
\end{equation}
we have proved our claim.

Now, consider the case when $M \geq 3$ and $m = 2,...,M-1$. We can
see that if $m < (M+1)/2$, $\gamma_{M+1-m}(\beta)$ is strictly
decreasing because the numerator and the denominator in
(\ref{eq:gamdef}) are strictly decreasing and strictly increasing
respectively as functions of $\beta$. Moreover, we can show the fact
that $\gamma_m(\beta) + \gamma_{M+1-m}(\beta)$ is strictly
decreasing, which can lead to the consequence that
$\gamma_{\frac{M+1}{2}}(\beta)$ is strictly decreasing for odd $M$
and thus $\gamma_{M+1-m}(\beta)$ is strictly decreasing for $m =
(M+1)/2$ as well. It suffices to prove that
\begin{equation}
\gamma\,'_m(\beta) + \gamma\,'_{M+1-m}(\beta) < 0
\label{eq:midgammaineq}
\end{equation}
for $\beta \in \left(0, \frac{\pi}{M}\right)$.
From (\ref{eq:gamma_m_sym}),
\begin{eqnarray}
\gamma_m(\beta) + \gamma_{M+1-m}(\beta) 
&=& \frac{1}{M} \;
\frac{\sin^2\left(M\phi_m(\beta)/2\right)}{\sin^2\left(
\phi_m(\beta)/2\right)} + \frac{1}{M} \;
\frac{\sin^2\left(M\phi_m(-\beta)/2\right)}{\sin^2\left(
\phi_m(-\beta)/2\right)} \nonumber \\
&=& \frac{1}{M} \cos^2 \left(\frac{M}{2}\beta\right)
\left[\csc^2\left( \phi_m(\beta)/2\right) + \csc^2\left(
\phi_m(-\beta)/2\right)\right],
\end{eqnarray}
because it can be shown that
\begin{equation}
\sin^2\left(M\phi_m(\beta)/2\right) =
\sin^2\left(M\phi_m(-\beta)/2\right) = \cos^2
\left(\frac{M}{2}\beta\right).
\end{equation}
By defining
\begin{equation}
f(\beta) \triangleq \cos^2 \left(\frac{M}{2}\beta\right),
\end{equation}
\begin{equation}
g_1(\beta) \triangleq \csc^2\left( \phi_m(\beta)/2\right), \quad
g_2(\beta) \triangleq \csc^2\left( \phi_m(-\beta)/2\right),
\end{equation}
we have the expression
\begin{equation}
\gamma\,'_m + \gamma\,'_{M+1-m} = \frac{1}{M} f\:(g_1+g_2)
\left(\frac{f'}{f}+\frac{g_1'+g_2'}{g_1+g_2}\right).
\end{equation}
Since $f>0$ and $g_1+g_2 > 0$ for $\beta \in \left(0,
\frac{\pi}{M}\right)$, we only need to show
\begin{equation}
h \triangleq \frac{f'}{f}+\frac{g_1'+g_2'}{g_1+g_2} < 0.
\label{eq:negah}
\end{equation}
Simple derivations give us
\begin{equation}
\frac{f'}{f} = -M \tan\left(\frac{M}{2}\beta\right),
\end{equation}
\begin{equation}
g_1' =  \frac{d}{d\beta} \left[ \csc^2\left(
\frac{\phi_m(\beta)}{2}\right) \right] = \csc^2\left(
\frac{\phi_m(\beta)}{2}\right) \cot\left(
\frac{\phi_m(\beta)}{2}\right),
\end{equation}
and similarly
\begin{equation}
g_2' =  -\csc^2\left( \frac{\phi_m(-\beta)}{2}\right) \cot\left(
\frac{\phi_m(-\beta)}{2}\right).
\end{equation}
We get
\begin{equation}
h(\beta) = -M \tan\left(\frac{M}{2}\beta\right) + \frac{\csc^2\left(
\frac{\phi_m(\beta)}{2}\right) \cot\left(
\frac{\phi_m(\beta)}{2}\right) - \csc^2\left(
\frac{\phi_m(-\beta)}{2}\right) \cot\left(
\frac{\phi_m(-\beta)}{2}\right)}{\csc^2\left(
\frac{\phi_m(\beta)}{2}\right)+\csc^2\left(
\frac{\phi_m(-\beta)}{2}\right)}.
\end{equation}
Because $0 < \phi_m(\beta)/2, \phi_m(-\beta)/2 < \pi$, applying the
mean value theorem yields
\begin{equation}
\frac{\csc^2\left( \frac{\phi_m(\beta)}{2}\right) \cot\left(
\frac{\phi_m(\beta)}{2}\right) - \csc^2\left(
\frac{\phi_m(-\beta)}{2}\right) \cot\left(
\frac{\phi_m(-\beta)}{2}\right)}{-\beta} = -2\csc^2\psi \cot^2\psi -
\csc^4\psi.
\end{equation}
for some $\psi \in \left(\phi_m(\beta)/2,\phi_m(-\beta)/2\right)$.
Then,
\begin{equation}
h(\beta) 
= -M \tan\left(\frac{M}{2}\beta\right) + \frac{\beta\csc^4\psi(2
\cos^2\psi + 1)}{\csc^2\left(
\frac{\phi_m(\beta)}{2}\right)+\csc^2\left(
\frac{\phi_m(-\beta)}{2}\right)}.
\end{equation}
We can see that for $\psi \in
\left[\phi_m(\beta)/2,\phi_m(-\beta)/2\right]$, $\csc^4\psi(2
\cos^2\psi + 1)$ has maximum at either $\psi = \phi_m(\beta)/2$ or
$\psi = \phi_m(-\beta)/2$ and let it be denoted by $\psi_b$. We are
ready to show the following series of inequalities
\begin{eqnarray}
h(\beta) &<& -M \tan\left(\frac{M}{2}\beta\right) +
\frac{\beta\csc^4\psi_b(2 \cos^2\psi_b + 1)}{\csc^2
\psi_b +1} \nonumber \\
&<&  -\frac{M^2}{2} \beta + \beta\csc^2\psi_b
\left(-2+\frac{10}{3-\cos 2\psi_b}\right) \nonumber \\
&<&  -\frac{M^2}{2} \beta + 3\beta\csc^2\psi_b
\nonumber \\
&<&  -\frac{M^2}{2} \beta + 3\beta\frac{1}{\sin^2\frac{\pi}{M}}\nonumber \\
&<&  
\frac{M^2}{2} \beta \left[-1 + \frac{6}{\pi^2\left\{1 -
\frac{1}{6}\left(\frac{\pi}{M}\right)^2\right\}^2}\right] \nonumber
\\
&<& 0,
\end{eqnarray}
where the last inequality holds as $M \geq 3$. This proves
(\ref{eq:negah}), and thus (\ref{eq:midgammaineq}) follows.

It is clear that $\sum_{i=1}^{m} \gamma_{(i)}(\beta)$ is strictly
decreasing for all $m \in \{1,...,M-1\}$ because
\begin{equation}
\gamma_{M+1-m}(\beta) \label{eq:decgamsingle}
\end{equation}
and
\begin{equation}
\gamma_m(\beta) + \gamma_{M+1-m}(\beta) \label{eq:decgampair}
\end{equation}
are strictly decreasing for $m=1,...,\lfloor\frac{M+1}{2}\rfloor$,
which we has been proved above, and $\sum_{i=1}^{m}
\gamma_{(i)}(\beta)$ becomes either the sum of (\ref{eq:decgampair})
for multiple $m$ or the sum of (\ref{eq:decgamsingle}) and
(\ref{eq:decgampair}) for multiple $m$. The validity of
(\ref{eq:noncenineq}) completes our proof.
\endproof

\myproof{Proof of Theorem \ref{th:tight}:} 
All functions in (\ref{eq:uppertight}) and (\ref{eq:lowertight})
take the value $0$ if and only if $x=0$. Thus, we can assume $p \in
(0,1)$. To show (\ref{eq:uppertight}), define $x_1 \triangleq
Q_M^{-1} (p)$ and $x_2 \triangleq F_{(M)}^{-1}(p|\nu)$ and this
yields
\begin{equation}
Q_M (x_1) = Q_{M} (x_2) W^{M-1} (x_2) = p.
\end{equation}
Let us introduce a new variable $x_3$ to obtain upper bound for
$x_2$ given by
\begin{equation}
x_3 \triangleq Q_M^{-1} \left(\frac{p}{W^{M-1}(x_1)}\right) \geq x_2
\geq x_1.
\end{equation}
Let us show $x_3 \approx x_1$, and then $x_2 \approx x_1$ in
(\ref{eq:uppertight}) follows immediately. The value of $x_1$ can be
computed using Sankaran's approximation in \cite{san63}, where it
has been suggested that for a random variable $X$ with the cdf
$F_{\chi'^2} (x|n,\delta)$, $\left\{X-(n-1)/2\right\}^{1/2} -
\left\{\delta+(n-1)/2\right\}^{1/2}$ is approximately zero mean
Gaussian with unit variance and this approximation improves if
either $n$ or $\delta$ increases. Thus as $M$ increases,
\begin{eqnarray}
x_1 &\approx& \frac{1}{2(K+1)}\left[\frac{1}{2} +
\left\{\left(2KM+\frac{1}{2}\right)^{\frac{1}{2}} + \Phi^{-1}(p)
\right\}^2\right] \nonumber \\
&\sim& \frac{K}{K+1} M, \label{eq:x1apprx}
\end{eqnarray}
where $\Phi^{-1}$ is the inverse function of the Gaussian cdf given
by
\begin{equation}
\Phi(x) = \frac{1}{\sqrt{2\pi}} \int_{-\infty}^{x}
e^{-\frac{t^2}{2}} dt.
\end{equation}
Let us use the notations $\mu_{F}$ and $\sigma^2_{F}$ to denote the
mean and variance of distribution $F$, respectively. Then, it can be
shown that
\begin{equation}
\mu_{W^{M-1}} \approx \frac{q_{M-1}+\zeta}{2(K+1)}, \quad
\sigma^2_{W^{M-1}} \approx \frac{\pi^2/6}{\left\{2(K+1)\right\}^2},
\label{eq:meanvarWM-1}
\end{equation}
where $q_{M-1} \triangleq W^{-1}\left(1-1/(M-1)\right) \approx \ln
(M-1)$ and $\zeta \triangleq 0.5772...$ (Euler's constant)
\cite{bai07}. For $x_1 > \mu_{W^{M-1}}$ (this is true for all $M >
C$ for some $C$), applying one-sided Chebyshev's inequality in
\cite[p.~152]{feller71} yields
\begin{equation}
1-W^{M-1}(x_1) \leq
\frac{1}{1+\left(x_1-\mu_{W^{M-1}}\right)^2/\sigma^2_{W^{M-1}}},
\end{equation}
and thus
\begin{equation}
1 \leq \frac{1}{W^{M-1}(x_1)} \leq 1+\epsilon_{M},
\end{equation}
where
\begin{equation}
0 < \epsilon_{M} \triangleq
\frac{\sigma^2_{W^{M-1}}}{\left(x_1-\mu_{W^{M-1}}\right)^2} \sim
\frac{\pi^2}{24}\frac{1}{K^2 M^2} \label{eq:epsilonrate}
\end{equation}
by (\ref{eq:x1apprx}) and (\ref{eq:meanvarWM-1}) as $M$ increases.
We have
\begin{eqnarray}
\lefteqn{x_3 - x_1} \nonumber \\
& \leq & Q_M^{-1} \left(p(1+\epsilon_{M})\right) - Q_M^{-1}
\left(p\right) \nonumber \\
& \approx & \frac{1}{K+1}\left(2KM+\frac{1}{2}\right)^{\frac{1}{2}}
\left\{\Phi^{-1}(p(1+\epsilon_{M})) - \Phi^{-1}(p)\right\} \nonumber \\
& & + \frac{1}{2(K+1)}
\left[\left\{\Phi^{-1}(p(1+\epsilon_{M}))\right\}^2 -
\left\{\Phi^{-1}(p)\right\}^2 \right] \nonumber \\
& \approx & \frac{1}{K+1}\left(2KM+\frac{1}{2}\right)^{\frac{1}{2}}
\epsilon_{M} p \: \frac{1}{\Phi'(\Phi^{-1}(p(1+\varepsilon)))}, \quad \varepsilon \in (1,\epsilon_{M}) \nonumber \\
& \approx & 0, \label{eq:firstconv}
\end{eqnarray}
as $M$ increases, because $\left(2KM+1/2\right)^{1/2} \cdot
\epsilon_{M} \approx 0$ from (\ref{eq:epsilonrate}). Hence,
(\ref{eq:uppertight}) is proved. Now, (\ref{eq:lowertight}) can be
shown similarly. For any $p \in (0,1)$, let us define $x_4
\triangleq Q_{a_M}^{-1} (\sqrt{p})$ and $x_5$ as
\begin{equation}
Q_{a_M}^2 (x_5) \cdot Q_{b_M}^2 (x_5) \cdot W^{M-4} (x_5) = p
\end{equation}
From (\ref{eq:minbound}), defining $x_6$ yields
\begin{equation}
x_6 \triangleq Q_{a_M}^{-1} \left(\frac{\sqrt{p}}{Q_{b_M} (x_4)
W^{\frac{M-4}{2}}(x_4)}\right) \geq x_5 \geq F_{(M)}^{-1}(p|0) \geq
x_4.
\end{equation}
Assuming $x_6 \approx x_4$, we have $F_{(M)}^{-1}(p|0) \approx x_4$
in (\ref{eq:lowertight}). 
Now, as $M$ increases, it can be shown that
\begin{equation}
x_4 \sim \frac{K}{K+1} a_M 
\end{equation}
as above. Note that $Q_{b_M} W^{\frac{M-4}{2}}$ is the distribution
of the maximum of two independent random variables following
$Q_{b_M}$ and $W^{\frac{M-4}{2}}$. It can be easily proved that
\begin{eqnarray}
\mu_{Q_{b_M} W^{\frac{M-4}{2}}} & \leq & \mu_{Q_{b_M}} +
\mu_{W^{\frac{M-4}{2}}} \nonumber \\
& \approx & \frac{K b_M +1}{K+1} + \frac{q_{\frac{M-4}{2}}+\zeta}{2(K+1)} \nonumber \\
& \sim & \frac{K}{K+1} b_M
\end{eqnarray}
and
\begin{eqnarray}
\sigma^2_{Q_{b_M} W^{\frac{M-4}{2}}} & \leq & \sigma^2_{Q_{b_M}} +
\sigma^2_{W^{\frac{M-4}{2}}} + \left(\mu_{W^{\frac{M-4}{2}}}\right)^2 \nonumber \\
& \approx & \frac{K b_M +1}{(K+1)^2} +
\frac{\pi^2/6}{\left\{2(K+1)\right\}^2} +
\left(\frac{q_{\frac{M-4}{2}}+\zeta}{2(K+1)}\right)^2 \nonumber \\
& \sim & \frac{K}{(K+1)^2} b_M.
\end{eqnarray}
Once again using one-sided Chebyshev's inequality,
\begin{equation}
1 \leq \frac{1}{Q_{b_M}(x_4) W^{\frac{M-4}{2}}(x_4)} \leq
1+\epsilon'_{M},
\end{equation}
where
\begin{eqnarray}
0 < \epsilon'_{M} &\triangleq& \frac{\sigma^2_{Q_{b_M}
W^{\frac{M-4}{2}}}}{\left(x_4-\mu_{Q_{b_M}
W^{\frac{M-4}{2}}}\right)^2} \nonumber \\
&\sim& \frac{1}{K} \frac{1/2 - 4/\pi^2}{(8/\pi^2 - 1/2)^2}
\frac{1}{M} \nonumber \\
&=& (0.9820...) \times \frac{1}{KM},
\end{eqnarray}
As $M$ increases, this implies $\left(2K a_M+1/2\right)^{1/2} \cdot
\epsilon'_{M} \approx 0$, which leads us $x_6 \approx x_4$ as in
(\ref{eq:firstconv}).
\endproof

\myproof{Proof of Theorem \ref{th:meanconverg}:} 
Let us show (\ref{eq:firstmeanconv}), first. Let $X_F$ denote a
random variable following any distribution $F$. Obviously,
${\it\Gamma}_{(M)}(\nu)$ is stochastically larger than $X_{Q_M}$
from (\ref{eq:maxbound}). Using the idea of coupling
\cite[Sec.~9.2]{ross96}, define
\begin{equation}
{\it\Gamma}^*_{(M)}(\nu) \triangleq
F_{(M)}^{-1}\left(Q_M(X_{Q_M})|\nu\right).
\end{equation}
Then, ${\it\Gamma}_{(M)}(\nu)$ and ${\it\Gamma}^*_{(M)}(\nu)$ share
the same distribution but ${\it\Gamma}^*_{(M)}(\nu) \geq X_{Q_M}$
with probability $1$. By the mean value theorem, we have
\begin{eqnarray}
h({\it\Gamma}^*_{(M)}(\nu)) - h(X_{Q_M}) = h'(\varepsilon) \left[
{\it\Gamma}^*_{(M)}(\nu) - X_{Q_M} \right],
\end{eqnarray}
for some $\varepsilon \in \left(X_{Q_M},
{\it\Gamma}^*_{(M)}(\nu)\right)$. Using this,
\begin{eqnarray}
\lefteqn{\left| \mathbb{E} \left[ h({\it\Gamma}_{(M)}(\nu)) -
h(X_{Q_M}) \right] \right|} \nonumber \\
&=& \left| \mathbb{E} \left[
h({\it\Gamma}^*_{(M)}(\nu)) - h(X_{Q_M}) \right] \right| \nonumber \\
&\leq& \mathbb{E} \left[ \left| h({\it\Gamma}^*_{(M)}(\nu)) -
h(X_{Q_M}) \right| \right] \nonumber \\
&\leq& C \cdot \mathbb{E} \left[ \left|
{\it\Gamma}^*_{(M)}(\nu) - X_{Q_M} \right| \right] \nonumber \\
&=& C \cdot \left| \mathbb{E} \left[ {\it\Gamma}_{(M)}(\nu) -
X_{Q_M} \right] \right|,
\end{eqnarray}
where $|h'|$ is bounded by $C$. Now, let us show $\mathbb{E} \left[
{\it\Gamma}_{(M)}(\nu)\right] \approx \mathbb{E} \left[ X_{Q_M}
\right]$. For any $x \geq 0$,
\begin{eqnarray}
Q_{M} (x) &\geq& F_{(M)}(x|\nu) = Q_{M} (x) W^{M-1} (x) \nonumber \\
&\geq& \left[ Q_{M} (x) + W^{M-1} (x) - 1 \right]^+,
\end{eqnarray}
where $[\cdot]^+$ is defined as
\begin{equation}
[y]^+ \triangleq \left\{ \begin{array}{ll}
y & \mbox{if } y \geq 0, \\
0 & \mbox{if } y < 0.
\end{array} \right.
\end{equation}
As $Q_{M} + W^{M-1} - 1$ is an increasing and continuous function of
$[0,\infty)$ onto $[-1,1)$, there exists only one $\alpha \geq 0$
such that
\begin{equation}
Q_{M} (\alpha) + W^{M-1} (\alpha) - 1 = 0.
\end{equation}
Therefore,
\begin{eqnarray}
\lefteqn{0 \leq \mathbb{E} \left[ {\it\Gamma}_{(M)}(\nu) - X_{Q_M}
\right]}
\nonumber \\
&=& \int_0^{\infty} \left[\left(1 - F_{(M)}(x|\nu)\right) - \left(1
- Q_{M} (x)\right)\right] \mathrm{d}x \nonumber \\
&\leq& \int_0^{\infty} \left[  Q_{M} (x) - \left[ Q_{M} (x) +
W^{M-1} (x) - 1 \right]^+ \right] \mathrm{d}x \nonumber \\
&=& \int_0^{\alpha} Q_{M} (x) \, \mathrm{d}x +
\int_{\alpha}^{\infty} \left[1 - W^{M-1} (x)\right] \mathrm{d}x \nonumber \\
&\leq& \int_0^{\beta} Q_{M} (x) \, \mathrm{d}x +
\int_{\beta}^{\infty} \left[1 - W^{M-1} (x)\right] \mathrm{d}x
\label{eq:diffMaxbound}
\end{eqnarray}
for any $\beta \geq 0$ as (\ref{eq:diffMaxbound}) can be minimized
by choosing $\beta = \alpha$. Let us obtain the upper bound for the
first term of (\ref{eq:diffMaxbound}) using the Marcum Q-function
defined and bounded as
\begin{eqnarray}
\Psi (a,b) &\triangleq& \int_b^{\infty} x e^{(x^2+a^2)/2}
I_0(ax)\,\mathrm{d}x \nonumber \\
&\geq& 1-\frac{a}{a-b} \exp \left(-\frac{1}{2}(a-b)^2\right) \;\;
\mbox{if } a > b, \label{eq:marcumdef}
\end{eqnarray}
where $I_0(x)$ is the modified Bessel function of the first kind
with order zero \cite{simon98}. Using the connection between the
Rice distribution and the noncentral chi-square distribution with
two degrees of freedom, it can be shown that
\begin{eqnarray}
Q_M(x) &=& F_{\chi'^2} (2(K+1)x|2,2KM) \nonumber \\
&=& 1-\Psi \left(\sqrt{2KM},\sqrt{2(K+1)x}\right).
\label{eq:Q2marcum}
\end{eqnarray}
From (\ref{eq:marcumdef}) and (\ref{eq:Q2marcum}), the first term in
(\ref{eq:diffMaxbound}) is bounded as
\begin{eqnarray}
\lefteqn{\int_0^{\beta} Q_{M} (x) \, \mathrm{d}x \leq \beta Q_{M}
(\beta)}
\nonumber \\
&\leq& \frac{\beta}{1 - \sqrt{(K+1)\beta/(KM)}} \exp
\left[-KM\left(1 - \sqrt{\frac{(K+1)\beta}{KM}}\right)^2\right],
\label{eq:Qbound}
\end{eqnarray}
for $\beta < KM/(K+1)$. If we take $\beta$ such that
\begin{equation}
\lim_{M \rightarrow \infty} \frac{\beta}{M} < \frac{K}{K+1},
\end{equation}
then (\ref{eq:Qbound}) goes to zero. Consider the second term of
(\ref{eq:diffMaxbound}). Note that $W$
is the exponential distribution, 
which has an increasing failure rate (IFR) \cite[Sec.~3.2]{bar75}.
From the chains of implication in \cite[p.~159]{bar75}, $W$ is a new
better than used (NBU) distribution, which is closed under the
formation of coherent systems including parallel systems, and thus
the distribution $W^{M-1}$ is a new better than used in expectation
(NBUE) as well as NBU. Using the bound for NBUE in
\cite[p.~187]{bar75}, the second term in (\ref{eq:diffMaxbound}) is
bounded as
\begin{equation}
\int_{\beta}^{\infty} \left[1 - W^{M-1} (x)\right] \mathrm{d}x \leq
\mu_{W^{M-1}} e^{-\beta/\mu_{W^{M-1}}}. \label{eq:Wbound}
\end{equation}
Note
\begin{equation}
\mu_{W^{M-1}} \approx \frac{\ln (M-1)+\zeta}{2(K+1)}
\end{equation}
from (\ref{eq:meanvarWM-1}), and thus we can find a sequence $\beta$
such that (\ref{eq:Wbound}) converges to zero as $M$ increases while
$\lim_{M \rightarrow \infty} \beta/M < K/(K+1)$, e.g., $\beta = 0.5
K\sqrt{M}/(K+1)$. We now prove (\ref{eq:secondmeanconv}). It can be
seen that
\begin{eqnarray}
Q^2_{a_M} (x) &\geq& F_{(M)}(x|0), \, Q^2_{a_M} (x) \, Q^2_{b_M}
(x) \nonumber \\
&\geq& Q^2_{a_M} (x) \, Q^2_{b_M} (x) \, W^{M-4} (x).
\end{eqnarray}
By the similar reasoning as above, it needs to be proved that
\begin{equation}
\mathbb{E} \left[ {\it\Gamma}_{(M)}(0)\right] \approx \mathbb{E}
\left[ X_{Q^2_{a_M}} \right] \label{eq:meanconv}
\end{equation}
as $M$ increases. We can easily show $\mathbb{E} \left[ X_{Q^2_{a_M}
Q^2_{b_M} W^{M-4}} \right] \approx \mathbb{E} \left[ X_{Q^2_{a_M}
Q^2_{b_M}} \right]$ as above. Assuming
\begin{equation}
\mathbb{E} \left[ X_{Q^2_{a_M} Q^2_{b_M}} \right] \approx \mathbb{E}
\left[ X_{Q^2_{a_M}} \right], \label{eq:meanconvaMbM}
\end{equation}
yields $\mathbb{E} \left[ X_{Q^2_{a_M} Q^2_{b_M} W^{M-4}} \right]
\approx \mathbb{E} \left[ X_{Q^2_{a_M}} \right]$, and thus
(\ref{eq:meanconv}) follows. Hence, we will show
(\ref{eq:meanconvaMbM}) to complete this proof. For this, we need to
find a sequence $\beta \geq 0$ such that
\begin{equation}
\int_0^{\beta} Q^2_{a_M} (x) \, \mathrm{d}x + \int_{\beta}^{\infty}
\left[1 - Q^2_{b_M} (x) \right] \mathrm{d}x \rightarrow 0
\label{eq:diffMinbound}
\end{equation}
as $M$ increases. To make the first term of (\ref{eq:diffMinbound})
diminish, $\beta$ can be chosen as
\begin{equation}
\lim_{M \rightarrow \infty} \frac{\beta}{a_M} < \frac{K}{K+1}.
\label{eq:betauppercond}
\end{equation}
As $Q_{b_M}$ is the noncentral chi-square distribution with two
degrees of freedom, $Q_{b_M}$ is IFR \cite{and08}, and thus
$Q^2_{b_M}$ is NBUE as above. From the definition of NBUE in
\cite[p.~159]{bar75}, the second term of (\ref{eq:diffMinbound}) is
upper bounded as
\begin{eqnarray}
\int_{\beta}^{\infty} \left[1 - Q^2_{b_M} (x)\right] \mathrm{d}x
&\leq& \mu_{Q^2_{b_M}} \left[1 - Q^2_{b_M} (\beta)\right] \nonumber \\
&\leq& 4 \mu_{Q_{b_M}} \left[1 - Q_{b_M} (\beta)\right].
\label{eq:QbMbound}
\end{eqnarray}
For $a < b$, Marcum Q-function is upper bounded as \cite{simon98}
\begin{equation}
\Psi (a,b) \leq \frac{b}{b-a} \exp \left(-\frac{1}{2}(b-a)^2\right).
\end{equation}
Then, (\ref{eq:QbMbound}) can be further bounded as
\begin{eqnarray}
\lefteqn{\int_{\beta}^{\infty} \left[1 - Q^2_{b_M} (x)\right]
\mathrm{d}x} \nonumber \\
&\leq& 4 \mu_{Q_{b_M}} \Psi
\left(\sqrt{2Kb_M},\sqrt{2(K+1)\beta}\right) \nonumber \\
&\leq& 4 \frac{Kb_M +1}{K+1} \frac{1}{1-1/\sqrt{(K+1)\beta/(Kb_M)}}
\exp \left[ -Kb_M \left(\sqrt{\frac{(K+1)\beta}{Kb_M}}-1\right)^2
\right],
\end{eqnarray}
which goes to zero if we take $\beta$ such that
\begin{equation}
\frac{K}{K+1} < \lim_{M \rightarrow \infty} \frac{\beta}{b_M} <
\infty. \label{eq:betalowercond}
\end{equation}
From the growth rates of $a_M$ and $b_M$ in (\ref{eq:aMgrowth}) and
(\ref{eq:bMgrowth}), $\beta$ can be selected such that
(\ref{eq:betauppercond}) and (\ref{eq:betalowercond}) are satisfied
simultaneously, e.g., $\beta = 0.25 \cdot KM/(K+1)$, which proves
(\ref{eq:meanconvaMbM}) and (\ref{eq:secondmeanconv}) consequently.

\endproof

\myproof{Proof of Lemma \ref{lm:capconverg}:} %
Obviously, $\log_2(1+\rho x)$ is integrable with respect to $Q_M$
and $Q^2_{a_M}$ as
\begin{equation}
\mathbb{E}\left[\log_2(1+\rho X_{Q_M})\right] \leq
\log_2\left(1+\rho \mu_{Q_M}\right)
\end{equation}
and
\begin{equation}
\mathbb{E}\left[\log_2(1+\rho X_{Q^2_{a_M}})\right] \leq
\log_2\left(1+\rho \mu_{Q^2_{a_M}}\right)
\end{equation}
by Jensen's inequality. From these and Theorem \ref{th:meanconverg},
we have
\begin{equation}
\mathbb{E} \left[ \log_2\left(1+\rho {\it\Gamma}_{(M)}(\nu)\right)
\right] \approx \mathbb{E}\left[\log_2(1+\rho X_{Q_M})\right]
\end{equation}
and
\begin{equation}
\mathbb{E} \left[ \log_2\left(1+\rho {\it\Gamma}_{(M)}(0)\right)
\right] \approx \mathbb{E}\left[\log_2(1+\rho X_{Q^2_{a_M}})\right]
\end{equation}
as $M$ increases. Then, we need to show that
\begin{equation}
\mathbb{E}\left[\log_2(1+\rho X_{Q_M})\right] \approx
\log_2\left(1+\rho \mu_{Q_M}\right) \label{eq:logQMmeanconv}
\end{equation}
and
\begin{equation}
\mathbb{E}\left[\log_2(1+\rho X_{Q^2_{a_M}})\right] \approx
\log_2\left(1+\rho \mu_{Q^2_{a_M}}\right),
\label{eq:logQsqaMmeanconv}
\end{equation}
as $M$ increases. Assuming that these are true,
(\ref{eq:maxergcapconv}) and (\ref{eq:minergcapconv}) follow
naturally from (\ref{eq:meanQM}) and (\ref{eq:meanQsqaM}). We will
now prove (\ref{eq:logQMmeanconv}). By Chebyshev's inequality, for
any given $\varepsilon > 0$, we have
\begin{eqnarray}
\lefteqn{\mathbb{E}\left[\log_2(1+\rho X_{Q_M})\right]} \nonumber \\
&=& \int_0^{\infty} \log_2(1+\rho x)\,\mathrm{d}Q_M(x) \nonumber \\
&\geq& \left[ \log_2(1+\rho \mu_{Q_M}) - \frac{\varepsilon}{2}
\right] \left[1 - \Pr\left\{\log_2\left(\frac{1+\rho X_{Q_M}}{1+\rho
\mu_{Q_M}}\right) \leq - \frac{\varepsilon}{2} \right\} \right] \nonumber \\
&=& \left[ \log_2(1+\rho \mu_{Q_M}) - \frac{\varepsilon}{2} \right]
\left[1 - \Pr\left\{\frac{X_M-\mu_{Q_M}}{\sigma_{Q_M}} \leq
-\frac{\mu_{Q_M}+1/\rho}{\sigma_{Q_M}}\left(1-2^{-\varepsilon/2}\right) \right\} \right] \nonumber \\
&\geq& \left[ \log_2(1+\rho \mu_{Q_M}) - \frac{\varepsilon}{2}
\right]
\left[1-\frac{\sigma_{Q_M}}{(1-2^{-\varepsilon/2})(\mu_{Q_M}+1/\rho)}
\right] \nonumber \\
&\geq& \log_2(1+\rho \mu_{Q_M}) - \frac{\varepsilon}{2} -
\log_2(1+\rho \mu_{Q_M})
\frac{\sigma_{Q_M}}{(1-2^{-\varepsilon/2})(\mu_{Q_M}+1/\rho)}
\nonumber \\
&\geq& \log_2(1+\rho \mu_{Q_M}) - \varepsilon
\end{eqnarray}
for large enough $M$ because $\mu_{Q_M} = (KM+1)/(K+1)$ given in
(\ref{eq:meanQM}) and $\sigma_M = \sqrt{KM+1}/(K+1)$, which proves
(\ref{eq:logQMmeanconv}). Moreover, (\ref{eq:logQsqaMmeanconv}) can
be shown similarly as $\mu_{Q^2_{a_M}} \geq \mu_{Q_{a_M}} =
(Ka_M+1)/(K+1)$ and $\sigma_{Q^2_{a_M}} \leq \sqrt{2}
\sigma_{Q_{a_M}} = \sqrt{2(2Ka_M+1)}/(K+1)$ by the variance bound in
\cite[p.~69]{david03}
\endproof

\bibliographystyle{IEEETran}
\bibliography{reference_list}

\end{document}